\documentclass{llncs}

\usepackage{amsmath}
\usepackage{graphicx}
\usepackage{color}

\begin{document}


\newcommand{\ket}[1]{| #1 \rangle}
\newcommand{\bra}[1]{\langle #1 |}
\newcommand{\braket}[2]{\langle #1 | #2 \rangle}

\newcommand{\comment}[1]{}

\newtheorem{Theorem}{Theorem}
\newtheorem{Lemma}{Lemma}
\newtheorem{Claim}{Claim}
\newtheorem{Corollary}{Corollary}

\def\U{\Uparrow}
\def\D{\Downarrow}
\def\L{\Leftarrow}
\def\R{\Rightarrow}


\title{Exceptional configurations of quantum walks with Grover's coin}

\author{Nikolajs Nahimovs, Alexander Rivosh
\thanks{NN is supported by EU FP7 project QALGO, AR is supported by ERC project MQC.}
}
\institute{Faculty of Computing, University of Latvia, Raina bulv. 19, Riga, LV-1586, Latvia.} 

\maketitle


\textbf{Abstract}.

We study search by quantum walk on a two-dimensional grid using the algorithm of Ambainis, Kempe and Rivosh \cite{AKR05}.
We show what the most natural coin transformation --- Grover's diffusion transformation --- has a wide class of exceptional configurations of marked locations, for which the probability of finding any of the marked locations does not grow over time. This extends the class of known exceptional configurations; until now the only known such configuration was the ``diagonal construction'' by \cite{AR08}.


\section{Introduction}

Quantum walks are the quantum counterparts of classical random walks \cite{Por13}. 
They have been useful for designing quantum algorithms for a variety of problems \cite{CC+03,AKR05,MSS05,BS06,Amb07}. In many of those applications, quantum walks are used as a tool for search.

To solve a search problem using quantum walks, we introduce the notion of marked locations. Marked locations correspond to elements of the search space that we want to find. We then perform a quantum walk on the search space with one transition rule at the unmarked locations, and another transition rule at the marked locations. If this process is set up properly, it leads to a quantum state in which marked locations have higher probability than the unmarked ones. This state can then be measured, finding a marked location with a sufficiently high probability. This method of search using quantum walks was first introduced in \cite{SKW03} and has been used many times since then.

We study search by quantum walk on a finite two-dimensional grid using the algorithm of Ambainis, Kempe and Rivosh (AKR). The original \cite{AKR05} paper proves that after $O(\sqrt{N \log{N}})$ steps, a quantum walk with one or two marked locations reaches a state that is significantly different from the initial state. Szegedy \cite{Sze04} has generalized this to an arbitrary number of marked locations. Thus, quantum walks can detect the presence of an arbitrary number of marked locations.
\cite{AKR05} also shows that for one or two marked locations, the probability of finding a marked location after $O(\sqrt{N \log{N}})$ steps is $O(1/\log{N})$. Thus, for one or two marked locations, the AKR algorithm can also find a marked location. For a larger number of marked locations, this is not always the case.
Ambainis and Rivosh \cite{AR08} have found an exceptional configuration of marked locations for which AKR algorithm fails to find any of marked locations.

A step of the AKR algorithm consists of two transformations: the coin-flip transformation, which acts on internal state of the walker and rearranges the amplitudes of going to adjacent locations, and the shift transformation, which moves the walker between the adjacent locations.
The original AKR algorithm uses $D$ --– Grover's diffusion transformation --– as the coin transformation for the unmarked locations and $-I$ as the coin transformation for the marked locations\footnote{According to authors of \cite{AKR05}, this coin transformation was chosen because it leads to a simpler proof.}.
Another natural choice for the coin transformation is $D$ for the unmarked locations and $-D$ for the marked locations.

Nahimovs and Rivosh \cite{NR15} has analysed the dependence of the running time of the AKR algorithm on the number and placement of marked locations and showed that the algorithm is inefficient for grouped marked locations (multiple marked locations placed near-by).
They showed that for a $k \times k$ group of marked locations, the AKR algorithm needs the same number of steps and has the same probability to find a marked location as for $4(k-1)$ marked locations placed as the perimeter of the group (with all internal locations being unmarked).
The reason for the inefficiency is the coin transformation used by the original AKR algorithm. The original coin transformation does not rearrange direction amplitudes within a marked location.
As a result, marked locations inside the group have almost no effect on the number of steps and the probability to find a marked location of the algorithm.

We try to solve the above problem by replacing the original coin transformation by one which rearranges amplitudes within a marked location.
We use the most natural choice of such coin transformation --- Grover's diffusion transformation. We show what while the modified algorithm works well for some of the ``problematic'' configurations, it has a wide class of exceptional configurations of marked locations, for which the probability to find any of marked locations does not grow over time. Namely, we prove that any block of marked locations of size $2m \times l$ or $m \times 2l$, that is the block with one of its sides consisting of even number of marked locations, is the exceptional configuration.
This extends the class of known exceptional configurations; until now the only known such configuration was the ``diagonal construction'' by \cite{AR08}.

\comment{
The paper is organized as follows. In the Section \ref{sec:Quantum_walks} we describe AKR quantum walk algorithm for two-dimensional grid. In the Section \ref{sec:Grovers_coin} we analyse AKR algorithm with Grover's coin and describe the class of exceptional configurations.
Section \ref{sec:Conclusions} give conclusions.
}
The AKR algorithm is very generic and can be adapted to other types of graphs. In the appendix we describe the AKR algorithm for general graphs and generalize the exceptional configurations that have been found.


\section{Quantum walks in two dimensions}
\label{sec:Quantum_walks}

Suppose we have $N$ items arranged on a two dimensional grid of size $\sqrt{N} \times \sqrt{N}$. We denote $n=\sqrt{N}$.
The locations on the grid are labelled by their $x$ and $y$ coordinates as $(x,y)$ for $x,y \in \{ 0, \dots, n-1\}$. We assume that the grid has periodic boundary conditions. For example, going right from a location $(n-1, y)$ on the right edge of the grid leads to the location $(0, y)$ on the left edge of the grid.

To introduce a quantum version of a random walk, we define a location register with basis states $\ket{i,j}$ for $i,j \in \{0,\dots,n-1\}$. Additionally, to allow non-trivial walks, we define a direction or coin register with four basis states, one for each direction: $\ket{\U}$, $\ket{\D}$, $\ket{\L}$ and $\ket{\R}$. Thus, the basis states of the quantum walk are $\ket{i,j,d}$ for $i,j \in \{0,\dots,n-1\}$ and $d \in \{\U,\D,\L,\R\}$. The state of the quantum walk is given by:
\\
$$
\ket{\psi(t)} = \sum_{i,j} (
\alpha_{i,j,\U}\ket{i,j,\U} + \alpha_{i,j,\D}\ket{i,j,\D} + 
\alpha_{i,j,\L}\ket{i,j,\L} + \alpha_{i,j,\R}\ket{i,j,\R} ).
$$

A step of the quantum walk is performed by first applying $I \otimes C$, where $C$ is unitary transform on the coin register. The most often used transformation on the coin register is the Grover's diffusion transformation $D$:

$$
D = \frac{1}{2} \left( 
\begin{array}{cccc}
-1 & 1 & 1 & 1 \\
1 & -1 & 1 & 1 \\
1 & 1 & -1 & 1 \\
1 & 1 & 1 & -1 
\end{array} \right).
$$
\\
Then, we apply the shift transformation $S$:
\\
$$
\begin{array}{lcl}
\ket{i,j,\U} & \rightarrow & \ket{i,j-1,\D} \\
\ket{i,j,\D} & \rightarrow & \ket{i,j+1,\U} \\
\ket{i,j,\L} & \rightarrow & \ket{i-1,j,\R} \\
\ket{i,j,\R} & \rightarrow & \ket{i+1,j,\L}
\end{array}
$$
\\
Notice that after moving to an adjacent location we change the value of the direction register to the opposite.
This is necessary for the quantum walk algorithm of \cite{AKR05} to work.

We start the quantum walk in the state 
$$
\ket{\psi_0} = \frac{1}{\sqrt{4N}} \sum_{i,j} \big( \ket{i,j,\U} + \ket{i,j,\D} + \ket{i,j,\L} + \ket{i,j,\R} \big).
$$
\\
It can be easily verified that the state of the walk stays unchanged, regardless of the number of steps. 

To use the quantum walk as a tool for search, we mark some locations. For the unmarked locations, we apply the same transformations as above. For the marked locations, we apply $-I$ instead of $D$ as the coin flip transformation. The shift transformation remains the same for both the marked and the unmarked locations.

Another way to look at the step of the algorithm is that we first perform a query $Q$ transformation, which flips signs of amplitudes of marked locations, then conditionally perform the coin transformation ($I$ or $D$ depending on whether the location is marked or not) and then perform the shift transformation $S$.
In the case of the modified coin ($D$ for unmarked locations and $-D$ for marked locations), the step of the algorithm consists of the query $Q$ followed by $D$ followed by $S$.

If there are marked locations, the state of the algorithm starts to deviate from $\ket{\psi(0)}$. It has been shown \cite{AKR05} that after $O(\sqrt{N\log{N}})$ steps, the inner product $\braket{\psi(t)}{\psi(0)}$ becomes close to $0$.

In the case of one or two marked locations, the AKR algorithm finds a marked location with $O(1 / \log{N})$ probability. 
The probability is small, thus, the algorithm uses amplitude amplification to get $\Theta(1)$ probability. The amplitude amplification adds an additional $O(\sqrt{\log{N}})$ factor to the number of steps. Thus, the total running time of the algorithm is $O(\sqrt{N} \log{N})$.


\section{Quantum walks with Grover's coin}
\label{sec:Grovers_coin}

The coin transformation used by the AKR algorithm does not rearrange amplitudes within a marked location. As it was shown in \cite{NR15}, this results in the AKR algorithm being inefficient for grouped marked locations.

In this section we consider an alternative coin transformation which rearranges amplitudes at both the marked and unmarked locations. As the most natural choice of such transformation we use $D$ and $-D$ as coin for the unmarked and marked locations, respectively. We refer this coin transformation as Grover's coin and the original coin transformation of the AKR algorithm as the AKR coin.

First, we compare the Grover and AKR coins for a $\sqrt{k} \times \sqrt{k}$ group of marked locations (``inefficient'' configuration of \cite{NR15}). We run a series of numerical experiments and demonstrate that in some cases, Grover's coin works better than AKR coin.

Next, we show a wide class of exceptional configurations of marked locations, for which the probability to find any of marked locations does not grow over time. We explain exceptional configurations based on stationary states of a step of the algorithm with Grover's coin.


\subsection{AKR vs Grover's coin: numerical experiment results.}

In this subsection, we compare the AKR algorithm with the Grover and AKR coins.
We consider $k$ marked locations placed as a $\sqrt{k} \times \sqrt{k}$ square and compare the number of steps and the probability to find a marked location for $\sqrt{k} \in [2,\dots,10]$ and grid sizes from $100 \times 100$ to $1000 \times 1000$ with step $100$. 

Table \ref{tbl:AKR_vs_Grover_k=9} shows the results of numerical simulations for $k = 9$ ($3 \times 3$ group of marked locations).
As one can see, the algorithm with Grover's coin needs more steps, however, it has much higher probability of finding a marked location and, thus, has smaller total running time (number of steps of the single run of the algorithm divided by square root of the probability).

\begin{table}[]
\vspace*{-3mm}
\centering
\begin{tabular}{|l|l|l|l|l|l|l|}
\hline
& \multicolumn{3}{l|}{AKR coin} & \multicolumn{3}{l|}{Grover's coin} \\ \hline
Grid size & Steps & Probability & Runtime & Steps   & Probability   & Runtime  \\ \hline
100       & 156   & 0.086454    & 531     & 318     & 0.556187      & 427      \\ \hline
200       & 345   & 0.066591    & 1337    & 653     & 0.527665      & 899      \\ \hline
300       & 544   & 0.063212    & 2164    & 993     & 0.510679      & 1390     \\ \hline
400       & 749   & 0.058022    & 3110    & 1337    & 0.499213      & 1893     \\ \hline
500       & 959   & 0.055813    & 4060    & 1685    & 0.49053       & 2406     \\ \hline
600       & 1172  & 0.055086    & 4994    & 2034    & 0.483683      & 2925     \\ \hline
700       & 1389  & 0.052851    & 6042    & 2386    & 0.478038      & 3451     \\ \hline
800       & 1608  & 0.051962    & 7055    & 2739    & 0.473336      & 3982     \\ \hline
900       & 1829  & 0.049888    & 8189    & 3093    & 0.469256      & 4516     \\ \hline
1000      & 2052  & 0.049255    & 9246    & 3449    & 0.465662      & 5055     \\ \hline
\end{tabular}
\vspace*{2mm}
\caption{Number of steps, probability and running time for the algorithm with the AKR and Grover coins for $k = 9$ and different $N$.}
\label{tbl:AKR_vs_Grover_k=9}
\vspace*{-7mm}
\end{table} 

Table \ref{tbl:AKR_vs_Grover_k=9_runtime} shows the ratio between running times of the algorithm with the AKR and Grover coins for $k=9$.
Table \ref{tbl:AKR_vs_Grover_runtime} shows the ratio for different $k$ and $N$.
As one can see, the ratio between the running times decreases with $k$ and increases with $N$.
The below results are obtained by running a series of  numerical simulations.
Thus, the interesting and important open question here is to find analytical formula giving the running time of the algorithm with AKR and Grover's coins for a group of marked locations.

\begin{table}[]
\centering
\begin{tabular}{|l|l|l|l|}
\hline
Grid size & AKR coin & Grover's coin & Ratio       \\  \hline
100       & 531      & 427           & 1.243559719 \\ \hline
200       & 1337     & 899           & 1.487208009 \\ \hline
300       & 2164     & 1390          & 1.556834532 \\ \hline
400       & 3110     & 1893          & 1.642894876 \\ \hline
500       & 4060     & 2406          & 1.687448047 \\ \hline
600       & 4994     & 2925          & 1.707350427 \\ \hline
700       & 6042     & 3451          & 1.75079687  \\ \hline
800       & 7055     & 3982          & 1.771722752 \\ \hline
900       & 8189     & 4516          & 1.813330381 \\ \hline
1000      & 9246     & 5055          & 1.829080119 \\ \hline
\end{tabular}
\vspace*{2mm}
\caption{Ratio between running times for the AKR and Grover coins for $k = 9$ and different $N$.}
\label{tbl:AKR_vs_Grover_k=9_runtime}
\vspace*{-7mm}
\end{table}

\begin{table}[]
\centering
\begin{tabular}{|l|l|l|l|l|}
\hline
Grid size & k = 9       & k = 25      & k = 49      & k = 81      \\ \hline
100       & 1.243559719 & 1.016453382 & 0.771014493 & 0.624553039 \\ \hline
200       & 1.487208009 & 1.286351472 & 1.059413028 & 0.829787234 \\ \hline
300       & 1.556834532 & 1.420867526 & 1.205965909 & 1.02627451  \\ \hline
400       & 1.642894876 & 1.480191554 & 1.268619838 & 1.123094959 \\ \hline
500       & 1.687448047 & 1.552176918 & 1.345050619 & 1.196541248 \\ \hline
600       & 1.707350427 & 1.631473534 & 1.406490777 & 1.224640497 \\ \hline
700       & 1.75079687  & 1.655281776 & 1.458191978 & 1.275856335 \\ \hline
800       & 1.771722752 & 1.695495113 & 1.500870777 & 1.344321812 \\ \hline
900       & 1.813330381 & 1.730009407 & 1.56015444  & 1.356277391 \\ \hline
1000      & 1.829080119 & 1.775771891 & 1.591205438 & 1.411492122 \\ \hline
\end{tabular}
\vspace*{2mm}
\caption{The ratio between the running times for the AKR and Grover coins for different $k$ and $N$.}
\label{tbl:AKR_vs_Grover_runtime}
\vspace*{-7mm}
\end{table}

For $k=4$, the quantum walk with Grover's coin does not find any of the marked locations. More precisely, the overlap between the current and initial state of the algorithm never reaches $0$, but stays close to $1$. Thus, the probability to find a marked location does not grow with the number of steps. The same holds for $k=16$, $k=36$, $k=64$, etc., that is, for any $k$ having even $\sqrt{k}$. Moreover, the same effect holds for any block of marked locations of size $2m \times l$ and $m \times 2l$, that is, the block with one of it sides consisting of an even number of marked locations. 

Therefore, while the algorithm with Grover's coin has a smaller running time, compared to the algorithm with the AKR coin, for some configurations, it completely fails for other configurations.


\subsection{Exceptional configurations of a quantum walk with Grover's coin.}

As it was mentioned in the previous subsection, the AKR algorithm using Grover's coin fails to find any group of marked locations of size $2m \times l$ or $m \times 2l$. In this subsection, we explain this phenomenon. First, we prove that a group of marked locations of size $1 \times 2$ (and by symmetry $2 \times 1$) is an exceptional configuration.
Next, we show how one can extend the argument to any group of size $2m \times l$ or $m \times 2l$.

\begin{figure}[ht]
\centering
\includegraphics[scale=0.5]{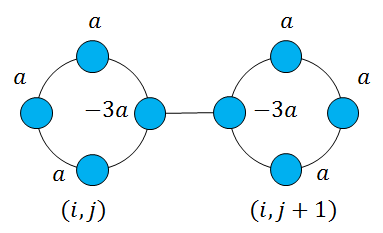}
\caption{Stationary state for $1 \times 2$ block of marked locations.}
\label{fig:1x2_stationary_state}
\end{figure}

Consider a grid of size $\sqrt{N} \times \sqrt{N}$ with two marked locations $(i,j)$ and $(i,j+1)$.
Let $\ket{\phi_{stat}^a}$ be a state having amplitudes of all basis states except $\ket{i,j,\R}$ and $\ket{i,j+1,\L}$ equal to $a$ and amplitudes of basis states $\ket{i,j,\R}$ and $\ket{i,j+1,\L}$ equal to $-3a$ (see figure \ref{fig:1x2_stationary_state}). 
Then this state is not changed by a step of the algorithm.

\begin{Theorem}
\label{thm:1x2_stationary_state}
Let locations $(i,j)$ and $(i,j+1)$ be marked and let
$$
\ket{\phi_{stat}^a} = \sum_{i,j,d}a\ket{i,j,d} - 4a\ket{i,j,\R} - 4a\ket{i,j+1,\L}.
$$
Then, $\ket{\phi_{stat}^a}$ is not changed by a step of the algorithm with Grover's coin. 
\end{Theorem}

\noindent
\textbf{Proof.}
Consider the effect of a step of the algorithm on $\ket{\phi_{stat}^a}$. The query transformation changes the signs of all the amplitudes of the marked locations. The coin transformation perform an inversion above the average: for non-marked locations, it does nothing as all amplitudes are equal to $a$; for marked locations, the average is $0$, so the inversion results in sign flip. Thus, $CQ$ does nothing for amplitudes of non-marked locations and twice flips the sign of amplitudes of marked locations. Therefore, we have $$CQ\ket{\phi_{stat}^a} = \ket{\phi_{stat}^a}.$$ 
The shift transformation swaps the amplitudes of near-by locations. For $\ket{\phi_{stat}^a}$, it swaps $a$ with $a$ and $-3a$ with $-3a$. Thus, we have $$SCQ\ket{\phi_{stat}^a} = \ket{\phi_{stat}^a}.$$
\qed

Consider the initial state of the algorithm
$$
\ket{\psi_0} = \frac{1}{\sqrt{4N}} \sum_{i,j} \big( \ket{i,j,\U} + \ket{i,j,\D} + \ket{i,j,\L} + \ket{i,j,\R} \big).
$$
It can be written as 
$$
\ket{\psi_0} = \ket{\phi_{stat}^a} + 4a(\ket{i,j,\R} + \ket{i,j+1,\L}),
$$ 
for $a=1/\sqrt{4N}$. Therefore, the only part of the initial state which is changed by the step of the algorithm is
$$
\sqrt{\frac{4}{N}}(\ket{i,j,\R} + \ket{i,j+1,\L}).
$$

Now, consider a group of marked locations of size $m \times 2l$.
It is equivalent to $m \times l$ groups of marked locations of size $1 \times 2$. Thus, by repeating the above construction $m \times l$ times, one can build the stationary state for the group. Moreover, if $m > 1$, then the group of size $2m \times l$ has multiple tilings by groups of size $2 \times 1$ and $1 \times 2$, where each tiling corresponds to a stationary state of the step of the algorithm.


\subsection{Alternative construction of stationary states.}

In this subsection we describe general conditions for a state to be a stationary state of the step of ARK algorithm with Grover's coin. 
and give an alternative construction of a stationary state for a group of marked locations.

\subsubsection{General conditions.}

A stationary state from the previous section has three properties:
\begin{itemize}
\item[1.] All directional amplitudes of unmarked locations are equal. This is necessary for the coin transformation to have no effect on the unmarked locations.
\item[2.] The sum of the directional amplitudes of any marked location is equal to $0$. This is necessary for the coin transformation to have no effect on marked locations.
\item[3.] Direction amplitudes of two adjacent locations pointing to each other are equal. This is necessary for the shift transformation to have no effect on the state.
\end{itemize}
It is easy to see that any state having these three properties is not changed by the step of AKR algorithm with Grover's coin and, thus, is a stationary state.

\subsubsection{Alternative construction of a stationary state.}

Consider a group of marked locations of size $m \times l$.
Without the loss of generality, let $m \leq l$.
We build the stationary state iteratively.
First, we set all directional amplitudes of the unmarked locations to $a$.
Next, we iteratively set amplitudes of the marked locations.
On each iteration we set the amplitudes of one rectangular layer of the marked locations, starting from the outer layer (the perimeter of the group).
The iteration is as follows:

\begin{itemize}
\item[1.] Set two directional amplitudes of a location pointing to its perimeter-wise neighbours to $-a$.
\item[2.] Set two other directional amplitudes of the location (pointing to the inner and the outer layers) to $a$.
\end{itemize}

\begin{figure}[ht]
\centering
\includegraphics[scale=0.5]{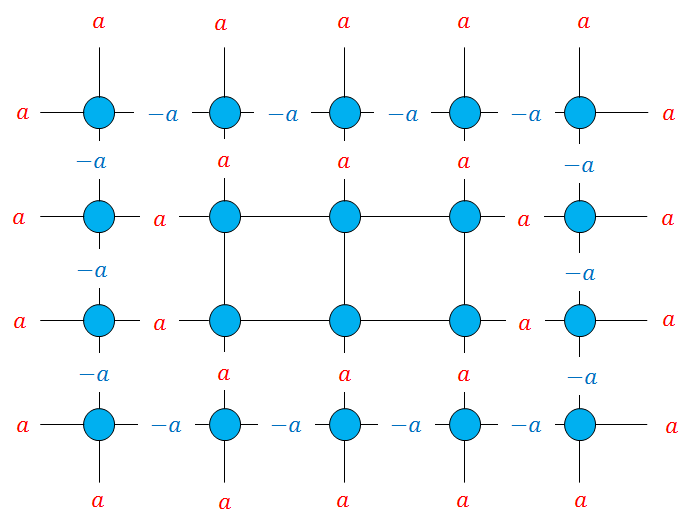}
\caption{The first iteration for a group of marked locations of size $4 \times 5$.}
\label{fig:4x5_stationary_state_first_step}
\end{figure}

\begin{figure}[ht]
\centering
\includegraphics[scale=0.5]{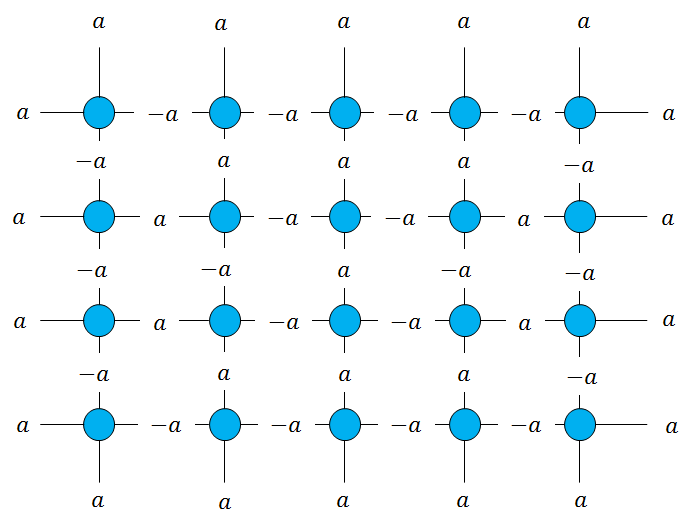}
\caption{The stationary state for a group of marked locations of size $4 \times 5$.}
\label{fig:4x5_stationary_state}
\end{figure}

\noindent
Figure \ref{fig:4x5_stationary_state_first_step} illustrates the first iteration of the construction for the group of marked locations of size $4 \times 5$. Amplitudes set on step 1 are colored in blue. Amplitudes set on step 2 are colored in red.
Figure \ref{fig:4x5_stationary_state} shows the resulting stationary state after all amplitudes are set.

\begin{figure}[ht]
\centering
\includegraphics[scale=0.5]{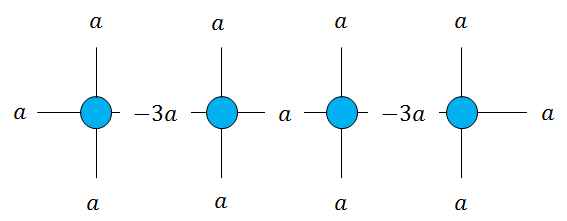}
\caption{The stationary state for a group of marked locations of size $4 \times 5$.}
\label{fig:1x4_stationary_state}
\end{figure}

The iteration reduces the size of the unprocessed group of marked locations from $m \times l$ to $m' \times l'$, where $m' = m-2$ and $l'=l-2$.
We repeat the iteration while $m' \geq 2$. 
If $m' = 0$, we have assigned values to all direction amplitudes and, thus, have built a stationary state. 
If $m' = 1$, there are three possibilities:

\begin{itemize}
\item $m' = l' = 1$. The construction is not possible. The initial group of marked locations was of odd-times-odd size.

\item $m' = 1$, $l' > 1$, $l$ is odd. The construction is not possible. The initial group  of marked locations was of odd-times-odd size.

\item $m' = 1$, $l' > 1$, $l'$ is even. Fill the remaining block by $1 \times 2$ constructions from Theorem \ref{thm:1x2_stationary_state} (figure \ref{fig:1x4_stationary_state} shows this for the block of marked locations of size $1 \times 4$.).
\end{itemize}

It is easy to see that for a group of marked locations of size odd-times-even and even-times-even, the above procedure leads to a state which satisfies all three properties of the stationary state.
First, all amplitudes of unmarked locations are equal to $a$. Second, the sum of amplitudes of a marked location is always $0$. Third, direction amplitudes of any two adjacent locations pointing to each other are equal.


\section{Conclusions and discussion}
\label{sec:Conclusions}

In this paper we have demonstrated a wide class of exceptional configurations for the AKR algorithm with Grover's coin. The above phenomenon is purely quantum. Classically, additional marked locations result in a decrease of the number of steps of the algorithm and an increase of the probability of finding a marked location.
Quantumly, as we have demonstrated in the paper, the addition of a marked location can drastically drop the probability of finding a marked location.

Another interesting consequence of the found phenomena is that the algorithm with Grover's coin ``distinguishes'' between odd-times-odd and even-times-even groups of marked locations. Moreover, if there are multiple odd-times-odd and even-times-even groups of marked locations, the algorithm will find only odd-times-odd groups and ``ignore'' even-times-even groups.
Nothing like this is possible for classical random walks without adding additional memory resources and complicating the algorithm.
The described phenomenon might have algorithmic applications which would be very interesting to find.



\newpage

\appendix

\section{General graphs}

In this appendix, we overview the stationary states of quantum walks with Grover's coin for general graphs.

\subsection*{Quantum walks on a general graph}

Consider a graph $G = (V, E)$ with a set of vertices $V$ and a set of edges $E$. Let $n = |V|$ and $m = |E|$. 
Let $N(x)$ be a neighbourhood of a vertex $x$, that is a set of vertices $x$ is adjacent to.
We define a location register with $n$ basis states $\ket{i}$ for $i \in \{1,\dots,n\}$ and a direction or coin register, which for a vertex $v_i$ has $d_i = \deg(v_i)$ basis states $\ket{j}$ for $j \in N(v_i)$.
The state of the quantum walk is given by:
\\
$$
\ket{\psi(t)} = \sum_{i=1}^{n} \sum_{j \in N(v_i)} \alpha_{i,j} \ket{i,j}.
$$

A step of the quantum walk is performed by first applying $I \otimes C$, where $C$ is a unitary transformation on the coin register. The usual choice of transformation on the coin register is Grover's diffusion transformation $D$.
Then, we apply the shift transformation $S$:
\\
$$
S = \sum_{i=1}^{n} \sum_{j \in N(v_i)} \ket{j,i} \bra{i,j} ,
$$
\\
which for each pair of connected vertices $i,j$ swaps an amplitude of vertex $i$ pointing to $j$ with an amplitude of vertex $j$ pointing to $i$.

We start the quantum walk in the equal superposition over all pairs vertex-direction:
$$
\ket{\psi_0} = \frac{1}{\sqrt{n \cdot \deg(G)}} 
\sum_{i=1}^{n} \sum_{j \in N(v_i)} \ket{i,j},
$$
where $\deg(G) = \sum_{i} \deg(v_i)$. 
It can be easily verified that the state of the walk stays unchanged, regardless of the number of steps. 

To use the quantum walk as a tool for search, we mark some vertices. For the unmarked vertices, we apply the same transformations as above. For the marked vertices, we apply $-I$ instead of $D$ as the coin flip transformation. The shift transformation remains the same for both the marked and unmarked vertices.

Another way to look at a step of the algorithm is that we first perform a query $Q$ transformation, which flips signs of amplitudes of marked vertices, then conditionally perform the coin transformation ($I$ or $D$ depending on whether a vertex is marked or not) and then perform the shift transformation $S$.
In case of the Grover's coin the step of the algorithm is the query $Q$ followed by $D$ followed by $S$.

\subsection*{Stationary states of the quantum walk with Grover's coin for general graphs}

\begin{figure}[ht]
\centering
\includegraphics[scale=0.5]{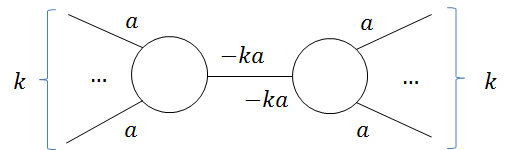}
\caption{Symmetric stationary state for $2$ marked vertices.}
\label{fig:2_vertex_symmetric_stationary_state}
\end{figure}

Consider a graph $G = (V, E)$ with two marked vertices $v_i$ and $v_j$.
Let vertices be connected and let each of them be connected to some other $k$ vertices.
Let $\ket{\phi_{stat}^a}$ be a state having amplitudes of all basis states except $\ket{i,j}$ and $\ket{j,i}$ equal to $a$ and amplitudes of basis states $\ket{i,j}$ and $\ket{j,i}$ equal to $-ka$ (see figure \ref{fig:2_vertex_symmetric_stationary_state}). 
Then this state is not changed by a step of the algorithm with Grover's coin.

\begin{Theorem}
\label{thm:2_vertex_stationary_state}
Let $G = (V,E)$ be a graph with two marked vertices $i$ and $j$;
let $(v_i,v_j) \in E$ and $N(v_i) = N(v_j) = k + 1$; and let
$$
\ket{\phi_{stat}^a} =
\sum_{i=1}^{n} \sum_{j \in N(v_i)} \ket{i,j} - 
(k+1)a (\ket{i,j} - \ket{j,i}).
$$
Then, $\ket{\phi_{stat}^a}$ is an eigenstate of a step of the quantum walk on $G$ with Grover's coin. 
\end{Theorem}

\noindent
\textbf{Proof.}
Consider the effect of a step of the algorithm on $\ket{\phi_{stat}^a}$. The query transformation changes the signs of all amplitudes of the marked vertices. The coin flip performs an inversion above the average: for unmarked vertices it does nothing as all amplitudes are equal to $a$; for marked vertices the average is $0$, so the inversion results in sign flip. Thus, $CQ$ does nothing for amplitudes of the unmarked vertices and twice flips the sign of amplitudes of the marked vertices. Therefore, we have $$CQ\ket{\phi_{stat}^a} = \ket{\phi_{stat}^a}.$$ 
The shift transformation swaps amplitudes of adjacent vertices. For $\ket{\phi_{stat}^a}$, it swaps $a$ with $a$ and $-ka$ with $-ka$. Thus, we have $$SCQ\ket{\phi_{stat}^a} = \ket{\phi_{stat}^a}.$$
\qed

The initial state of the algorithm $\ket{\psi_0}$ can be written as 
$$
\ket{\psi_0} = \phi_{stat}^a + (k+1)a(\ket{i,j} + \ket{j,i}),
$$ 
for $a=1/\sqrt{n \cdot \deg(G)}$. Therefore, the only part of the initial state, which is changed by a step of the algorithm, is
$$
\frac{k+1}{\sqrt{n \cdot \deg(G)}}(\ket{i,j} + \ket{j,i}).
$$

\begin{figure}[ht]
\centering
\includegraphics[scale=0.5]{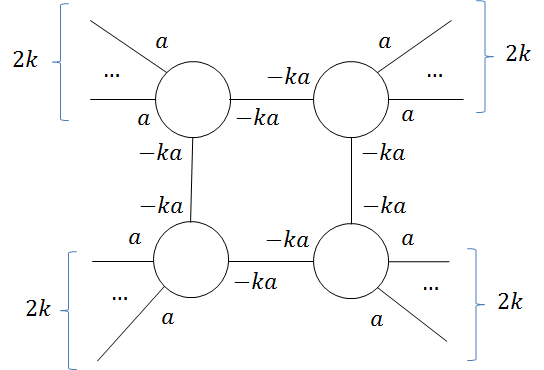}
\caption{Symmetric stationary state for $4$ marked vertices.}
\label{fig:4_vertex_symmetric_stationary_state}
\end{figure}

Next figures show similar constructions for three (figure \ref{fig:3_vertex_symmetric_stationary_state}) and four (figure \ref{fig:4_vertex_symmetric_stationary_state}) marked vertices. We give them without a proof (which is similar to the two marked vertex case). It is easy to see how one can extend the construction to any number of marked vertices.

\begin{figure}[ht]
\centering
\includegraphics[scale=0.5]{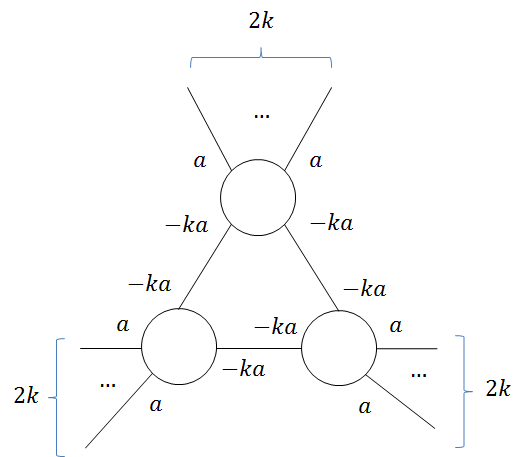}
\caption{Symmetric stationary state for $3$ marked vertices.}
\label{fig:3_vertex_symmetric_stationary_state}
\end{figure}

\begin{figure}[ht]
\centering
\includegraphics[scale=0.5]{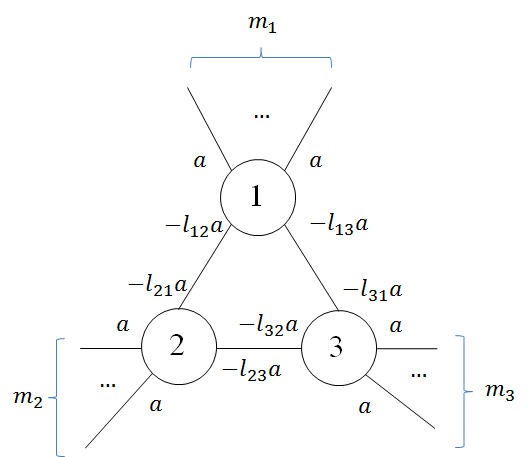}
\caption{Generic stationary state for $3$ marked vertices.}
\label{fig:3_vertex_generic_stationary_state}
\end{figure}

The above constructions are symmetric in the sense that each of the marked vertices has the same number of neighbours. One can also construct a stationary state without this restriction. The figure \ref{fig:3_vertex_generic_stationary_state} shows the general stationary state 
of three marked locations. The parameters of the construction (number of adjacent vertices) are restricted by Equation \ref{eq:3_vertex_generic_stationary_state}.

\begin{equation}
\label{eq:3_vertex_generic_stationary_state}
\begin{cases} 
l_{12} + l_{12} = m_1 \\ 
l_{21} + l_{23} = m_2 \\
l_{31} + l_{32} = m_3 \\
l_{12} = l_{21} \\
l_{23} = l_{32} \\
l_{31} = l_{13} 
\end{cases} 
.
\end{equation}
\\
For example, for $l_{12} = l_{21} = 1$, $l_{23} = l_{32} = 2$ and $l_{31} = l_{13} = 3$ we will have $m_1 = 4$, $m_2 = 3$ and $m_3 = 5$.

Again, it is easy to see how one can extend the construction to any number of marked vertices.


\end{document}